\documentclass[12pt]{article}
\usepackage[utf8]{inputenc} 
\usepackage[T1]{fontenc}    
\usepackage{hyperref}       
\usepackage{url}            
\usepackage{booktabs}       
\usepackage{amsfonts}       
\usepackage{amsmath}
\usepackage{graphicx}
\usepackage{enumitem}
\usepackage{indentfirst}
\usepackage{multirow}
\usepackage{libertine}
\usepackage{gensymb}
\usepackage[english]{babel}
\usepackage[top=2.5cm, bottom=2.5cm, left=2.5cm, right=2.5cm]{geometry}
\usepackage{natbib} 
\usepackage{caption}
\usepackage{subcaption}
\usepackage{xy}
\xyoption{all}
\usepackage{arydshln}

\DeclareMathOperator*{\argmax}{arg\,max}
\DeclareMathOperator*{\argmin}{arg\,min}

\usepackage{color}

\def\Chat{{\widehat C}} 
\def\Fhat{{\widehat F}} 
\def\hattheta{{\hat \theta}} 

\def\bfu{\boldsymbol{u}}
\def\bfx{\boldsymbol{x}}

\def\bftheta{\boldsymbol{\theta}}
\def\bfpsi{\boldsymbol{\psi}}

\def\hbftheta{{\widehat\bftheta}}

\title{Assessing Copula Models for Mixed Continuous-Ordinal Variables}
\author{
Shenyi Pan\thanks{Department of Statistics, University of British Columbia, Vancouver, BC Canada V6T 1Z4. Email: \href{mailto:shenyi.pan@stat.ubc.ca}{shenyi.pan@stat.ubc.ca}.}
\and
Harry Joe\thanks{Department of Statistics, University of British Columbia, Vancouver, BC Canada V6T 1Z4. Email: \href{mailto:harry.joe@ubc.ca}{Harry.Joe@ubc.ca}.}
}

\begin{document}
\maketitle

\begin{abstract}
Vine pair-copula constructions exist for a mix of continuous and ordinal variables. 
In some steps, this can involve estimating a bivariate copula for a pair
of mixed continuous-ordinal variables.
To assess the adequacy of copula fits for such a pair, diagnostic and visualization methods based on normal score plots and conditional Q-Q plots are proposed. The former utilizes a latent continuous variable for the ordinal variable. 
Using the Kullback-Leibler divergence, existing probability models for mixed continuous-ordinal variable pair are assessed for the adequacy of fit with simple parametric copula families. 
The effectiveness of the proposed visualization and diagnostic methods is illustrated on simulated and real datasets.
\end{abstract}

\noindent
Keywords: parametric copula, empirical beta copula,
Kullback-Leibler divergence,
location-scale mixture models, normal scores, ordinal regression, 
polyserial correlation.

\section{Introduction} \label{sec:intro}

Vine pair-copula constructions have been used for a mix of continuous and
discrete/ordinal variables in 
\cite{Stoeber.Hong.ea2015},
Section 3.9.5 of \cite{joe2014dependence}, and
\cite{chang2019prediction}.
The latter is concerned with the use of vine constructions for prediction models based on explanatory variables that are a mix of continuous and ordinal variables. 

In some steps of the vine construction, the estimation involves a bivariate copula for a pair of mixed continuous-ordinal variables.
The main objective of this paper is to assess the adequacy of the fit of parametric copula families for a pair of variables, one of which is continuous and the other is ordinal with a few categories. 
To decide on possible suitable bivariate parametric copula families, the two
variables can be visualized using normal score plots by converting the
ordinal variable into an appropriate latent continuous variable. 
After fitting candidate copula families, quantile-quantile (Q-Q) plots of the continuous variable conditioned on each category of the ordinal variable can be used to assess the adequacy of the fit.

To theoretically assess whether commonly used parametric copula families can fit mixed continuous-ordinal 
variables well, we consider some bivariate distributions or models proposed in the statistical literature for one ordinal and one continuous variable. 
The Kullback-Leibler (KL) divergence is used to assess the adequacy of copula-based approximations. 
For mixture models, we find that simple 1- or 2-parameter parametric copula families can lead to good approximations when there are homoscedastic and roughly equally spaced components and the number
of mixture components is small. 
For conditional probit and logit models, Gaussian or t copulas can provide perfect or near perfect matches when the continuous variable follows a normal distribution. 
Otherwise, simple parametric bivariate copula families can be inadequate. 
In particular, in mixture models, as the locations become more dispersed or when
there is heteroscedasticity, the best approximating simple parametric copula families 
based on the KL divergence can have reflection asymmetry, tail asymmetry, or tail dependence, but the fit can still be inadequate based on the conditional Q-Q plots.

When simple parametric copula families do not provide good fits, nonparametric copulas can be fitted with appropriate adaptations to an ordinal variable using the latent continuous variable. 
We indicate how to adapt the empirical beta copulas
(\cite{segers2017empirical} for use with a pair of mixed continuous and ordinal variables as an example.

The remainder of this paper is organized as follows.
Section~\ref{sec:copula} gives an overview of copulas when fitted to a pair of mixed continuous-ordinal variables. 
Section~\ref{sec:diagnostics} proposes visualization methods via normal score plots and copula model diagnostic methods via conditional Q-Q plots for a pair of mixed continuous-ordinal variables. 
Section~\ref{sec:kl} discusses approaches to assessing the adequacy of approximations by computing the KL divergence of a copula-based model from a given density.
Section~\ref{sec:fitting} covers procedures for fitting parametric and nonparametric copula models to mixed continuous-ordinal variables. 
The proposed visualization and diagnostic methods are illustrated on simulated datasets in Section~\ref{sec:illus} and demonstrated on a real dataset in Section~\ref{sec:app}. 
Section~\ref{sec:conc} has final discussions.

\section{Copulas for Mixed Continuous-Ordinal Variables} 
\label{sec:copula}

A copula is a multivariate distribution function with univariate Uniform $(0,1)$ margins. 
According to Sklar's theorem (\cite{sklar1959fonctions}), a $d$-variate distribution $F$ is a composition of a copula $C$ and its univariate marginal distributions $F_1,\ldots,F_d$; that is,
$F(\bfx)=C(F_1(x_1),\ldots,F_d(x_d))$, for $\bfx\in\mathbb{R}^d$.
If $F$ is a continuous $d$-variate distribution function with univariate margins $F_1, \ldots, F_d$, and quantile functions $F_1^{-1},\ldots,F_d^{-1}$, then the copula
$C(\bfu)=F(F_1^{-1}(u_1),\ldots,F_d^{-1}(u_d))$, for $\bfu\in[0,1]^d$,
is the unique choice. 
If $F$ is a $d$-variate distribution function of mixed continuous-ordinal variables, then the copula is only unique on the set $\text{Range}(F_1)\times\cdots\times\text{Range}(F_d)$. 
Hence, the copula associated with $F$ is non-unique when some variables are ordinal. 
Nevertheless, parametric copula families can still be used for data applications. 
Many non-uniqueness results are shown in \cite{Genest.Neslehova2007} for the completely discrete/ordinal case. 
For the bivariate case of one ordinal and one continuous variable, 
one candidate for the non-unique copula is based on
conditionally uniform given identities that the copula must satisfy, extending an idea on page 122 of \cite{joe2014dependence}. 
This corresponds to defining an appropriate latent continuous variable.

Consider an ordinal variable $X$ that can take $k$ distinct ordered values $\{v_1, \ldots, v_k\}$  and a continuous variable $Y$. 
Assume the labeling $v_j=j$ for $j=1,\ldots,k$ without loss of generality.
Let $F_{X,Y}$ be the cumulative distribution function (CDF) of $(X,Y)$, with copula $C_{X,Y}$
that is unique on $\{0,F_X(1),\ldots,F_X(k-1),1\}\times [0,1]$. 
Let $Z$ be a continuous latent variable associated with $X$ with distribution $F_Z$. 
If $Z = F^{-1}_Z(U_Z)$ and
\begin{equation}
  U_Z \sim U\left(F_X(i-1), F_X(i)\right), \text{ for } X = i, \quad i\in\{1,\ldots,k\},
  \label{eq: Z_cond}
\end{equation}
then it can be shown that $C_{Z,Y}(F_X(x),F_Y(y)) = C_{X,Y}(F_X(x),F_Y(y))$ holds
for $x\in\{1,\ldots,k\}$ and $F_Y(y)\in [0,1]$. 

The proof is as follows.
Note that $U_Z$ in \eqref{eq: Z_cond} has $U(0,1)$ distribution.
Hence, there is a unique copula $C_{Z,Y}$ for $(Z,Y)$ since $Z$ is continuous. 
Let $U_Y=F_Y(Y)$ and $U_X=F_X(X)$.
By Sklar's theorem, any copula $C_{X,Y}$ for $(X,Y)$ satisfies 
 $$C_{X,Y}\left(F_X(i),F_Y(y)\right) = \mathbb{P}\left(U_X \leq F_X(i), U_Y \leq F_Y(y)\right), \quad i=1,\ldots,k, \quad  -\infty<y<\infty. $$ 
For a given $i\in \{1,\ldots,k\}$, 
$U_X \leq F_X(i)$ implies that $X$ can take a value in $\{1, \ldots, i\}$. 
The corresponding $U_Z$ thus follows one of the uniform distributions $U(0, F_X(1)), \dots, U(F_X(i-1), F_X(i))$. 
This implies that $U_Z \leq F_X(i)$ also holds. 
Similarly, $U_Z \leq F_X(i)$ implies that $X\le i$ or $U_X \leq F_X(i)$. 
Hence, $U_Z \leq F_X(i)$ and $U_X \leq F_X(i)$ are equivalent events,
and 
\begin{eqnarray*}
  &&C_{Z,Y}\left(F_X(i),F_Y(y)\right)=
  \mathbb{P}\left(U_Z \leq F_X(i), U_Y \leq F_Y(y)\right)  \\
  &&= \mathbb{P}\left(U_X \leq F_X(i), U_Y \leq F_Y(y)\right)
  = C_{X,Y}\left(F_X(i),F_Y(y)\right).
\end{eqnarray*}
Thus $C_{Z,Y}$ matches $C_{X,Y}$ on $\{0,F_X(1),\ldots,F_X(k-1),1\}\times [0,1]$.

Note that the condition in \eqref{eq: Z_cond} only requires that $U_Z$ follows a piecewise uniform distribution given different values of $X$. 
However, there are no restrictions on the joint distribution of $U_Z$ and $U_Y$ given $U_X$.
Given different joint distributions of $[U_Z,U_Y|X=i]$,
the resulting copulas $C_{Z,Y}$ are also different. 
This shows the non-uniqueness of copulas for the case of one ordinal and one continuous variable. 
Examples of generating different sets of $U_Z$ values that satisfy the condition in \eqref{eq: Z_cond} are illustrated in the next section.

\section{Copula Model Diagnostics for Mixed Continuous-Ordinal} 
\label{sec:diagnostics}

There are diagnostic plots and measures of asymmetry or dependence
(Chapter 1 and Sections 2.12--2.15 of \cite{joe2014dependence}) to
assess whether 1-parameter or 2-parameter bivariate copula families
are useful for pairs of continuous variables with moderate to strong
dependence.
In this section, we discuss plots to help decide on possible
parametric copula families for a continuous-ordinal pair of variables, and assess the adequacy of fitted copulas
for a random sample with mixed continuous-ordinal variables. 
Visualization using normal score plots is presented in Section~\ref{subsec:normal_scores}. 
A diagnostic procedure for copula estimates using 
conditional Q-Q plots is given in Section~\ref{subsec:diag}. 

With an ordinal variable $X$ and a continuous variable $Y$,
suppose there is a random sample from $F_{X,Y}$ consisting of $n$ pairs $(x_i, y_i)$ for $i=1,\ldots,n$, with $x_i \in \{1, \ldots, k\}$. 
For each $j\in \{1, \ldots, k\}$, let $n_j$ be the cardinality of $\{i: x_i=j\}$.
Let $\Fhat_X$ be the  empirical distribution function of $\{x_i\}$.
For the continuous variable $Y$, a parametric or nonparametric model can be applied to estimate $F_Y$ with $\Fhat_Y$.
For the $i$th observation, let $u_{iX}^+ = \Fhat_X(x_i)$, $u_{iX}^- = \lim_{t\uparrow x_i}\Fhat_X(t) = \Fhat_X(x_i^-)$, and $u_{iY} = \Fhat_Y(y_i)$.

\subsection{Visualizations via Normal Score Plots}
\label{subsec:normal_scores}

For a pair of continuous variables, normal score plots are often used to visualize the copula for the variables and check for deviations from Gaussian dependence. 
With the ordinal variable $X$, $\{u_{iX}^+\}$ is a set of discrete values.
To visualize the relationship between $X$ and $Y$, we use 
an appropriate latent continuous variable associated with $X$.

For the normal score plots, 
we assume that the ordinal variable $X$ is generated from a latent standard normal variable $Z$ with estimated cutpoints $\zeta_{j} = \Phi^{-1}\left((n_1 + \cdots + n_j)/{n}\right)$ for $j = 1, \ldots, k-1$, where $\Phi$ and $\Phi^{-1}$ are the CDF and quantile function of the standard normal distribution.
Let $\zeta_{0} = -\infty$ and $\zeta_k=\infty$. 
We would like the latent variable $Z$ to be generated in such a way that the correlation between $Z$ and $Y$ is close to the correlation between $X$ and $Y$ in order for the visualization to preserve the strength of correlation from the original variables.

One measure of the association between an ordinal and a continuous variable is the polyserial correlation (see \cite{olsson1982polyserial} and Section 2.12.7 of \cite{joe2014dependence}). 
The polyserial correlation between $X$ and $Y$ is defined as
\begin{equation*}
\rho_N = \argmax_{\rho}\sum_{i=1}^{n}\log\left\{\phi\left(\Phi^{-1}(u_{iY})\right)\left[\Phi\left(\frac{\zeta_{x_i}-\rho \Phi^{-1}(u_{iY})}{\sqrt{1-\rho^2}}\right) - \Phi\left(\frac{\zeta_{x_i-1} - \rho \Phi^{-1}(u_{iY})}{\sqrt{1-\rho^2}}\right)\right]\right\},
\end{equation*}
where $\phi$ is the probability density function (PDF) of the standard normal distribution.

Let $\mathcal{S}_j = \{i: x_i=j\}$, $j = 1,\ldots,k$, be the set of indices of $X$ taking value $j$. 
We propose to generate the normal scores of the latent variable $Z$ for the ordinal variable $X$ in the following steps. 

\begin{enumerate}
\item Compute the polyserial correlation $\rho_N$ between $X$ and $Y$.
For $j = 1, \ldots, k$, perform steps 2 to 5 on $\mathcal{S}_j$:

\item 
For each $i\in \mathcal{S}_j$, independently generate $\omega_i \sim U(0,1)$. Let
$$ 
u_{iW} = \Phi\left(\sqrt{1-\rho_N^2}\Phi^{-1}(\omega_i) + \rho_N\Phi^{-1}(u_{iY})\right),
$$
assuming a bivariate Gaussian copula with correlation $\rho_N$.
Note that the generated $u_{iW}$ are in the range of $[0, 1]$. 
Let $\bfu_{W,j}=\{ u_{iW}: i\in \mathcal{S}_j\}$ be a vector
of length $n_j$.

\item
For each $i\in \mathcal{S}_j$, independently generate
$\psi_i\sim U(n^{-1}\sum_{t=1}^{j-1}n_t, n^{-1}\sum_{t=1}^{j}n_t)$ 
for $j > 1$, or $\psi_i\sim U(0,n_1/n)$ for $j = 1$. 
Let $\bfpsi_j=\{\psi_i:  i\in \mathcal{S}_j\}$ be a vector of length $n_j$.

\item For each $u_{iW}$ with $i\in \mathcal{S}_j$, find its rank in $\bfu_{W,j}$. 
Replace the value of $u_{iW}$ by $\psi_\ell$ that has the same rank in $\bfpsi_j$,
and denote this value by $u_{iZ}$. 
Let $\bfu_{Z,j}=\{u_{iZ}: i\in \mathcal{S}_j\}$.
\item Let $z_i = \Phi^{-1}(u_{iZ})$ for each $i\in\mathcal{S}_j$.
\end{enumerate}

The $\{u_{iW}\}$ generated in step 2 does not have a uniform distribution. 
The additional steps 3 and 4 match the ranks of $u_{iW}$ with a random vector generated from the uniform distribution. 
This procedure ensures that the generated normal scores $\{z_i\}$ fall into the desired bins separated by the cutpoints $\boldsymbol{\zeta}$ and marginally follow the standard normal distribution. 
Within each bin, 
$\{z_i\}$ and $\{\Phi^{-1}(u_{iY})\}$ also have similar correlation to $\rho_N$. 
The two sets of normal scores, $\{z_i: i=1,\ldots,n\}$ and 
$\{\Phi^{-1}(u_{iY}): i=1,\ldots,n\}$, can then be plotted against each other to decide on suitable parametric copula families. 
If a bivariate copula model can fit $\{(u_{iZ},u_{iY})\}$ well, then it can also potentially provide adequate fits to $\{(x_i,y_i)\}$.

Let $F_{n,X}(j)=\Fhat_X(j)$ for $j = 1,\dots,k$.
As $n\to\infty$, $F_{n,X}(j) = n^{-1}\sum_{t=1}^j n_t \to F_X(j)$ for $j = 1,\dots,k$. 
Therefore, each element $\psi_i$ of $\bfpsi_j$ in Step 3 of the proposed algorithm above follows $U\left(F_{n,X}(j-1), F_{n,X}(j)\right)$ and satisfies the condition~\eqref{eq: Z_cond} as $n\to\infty$. 
Similarly, each element of $\bfu_{Z,j}$ also satisfies the condition~\eqref{eq: Z_cond} as $n\to\infty$, since $\bfu_{Z,j}$ is a permutation of $\bfpsi_j$. 
However, the correlation for $\{(u_{iZ},u_{iY})\}$ is stronger than the correlation for $\{(\psi_i,u_{iY})\}$. 
The bivariate empirical copulas of the two sets $\{(\psi_i, u_{iY}): i = 1,\dots, n\}$ and
$\{(u_{iZ}, u_{iY}): i = 1,\dots, n\}$ evaluated at $(u_x,u_y)$ are the same for empirical CDF values $u_x=F_{n,X}(j)$ and arbitrary $u_y\in(0,1)$. 

\subsection{Diagnostics of Estimated Copula} 
\label{subsec:diag}

With $F_{XY}(x,y)=C_{XY}(F_X(x),F_Y(y))$ for a copula,
 the conditional distribution of the continuous variable $Y$ given $X=x$,
where $x$ is a possible value of the ordinal variable $X$, is
\begin{equation}
F_{Y|X}(y|x) = \frac {F_{XY}(x,y)-F_{XY}(x^-,y)} {F_X(x)-F_X(x^-)}
  =\frac {C_{XY}(F_X(x), F_Y(y)) - C_{XY}(F_X(x^-), F_Y(y))} {F_X(x)-F_X(x^-)}.
  \label{eq:condcdf}
\end{equation}

If $\Chat_{XY}= C(\cdot;\hbftheta)$ is an estimate from a parametric family,
let
$$\Fhat_{Y|X}(y|x) = 
  \frac {\Chat_{XY}(\Fhat_X(x), \Fhat_Y(y)) - \Chat_{XY}(\Fhat_X(x^-), \Fhat_Y(y))} {\Fhat_X(x)-\Fhat_X(x^-)}
$$
be the estimated conditional distribution.
Given a quantile level $q$, 
the conditional quantile $\Fhat^{-1}_{Y|X}(q|x)$ is obtained 
as $\Fhat^{-1}_Y(v)$ where  $v$ is the root to the equation 
 $$ \frac{\Chat_{XY}\bigl(\Fhat_X(x), v) -
 \Chat_{XY}\bigl(\Fhat_X(x^-), v\bigr)}{\Fhat_X(x) - \Fhat_X(x^-)} = q. $$

To assess the fit of $\Chat_{XY}$, conditional Q-Q plots can be generated based on
$\Fhat^{-1}_{Y|X}(\cdot|j)$ for $j=1,\ldots,k$.
For the $j$th conditional Q-Q plot, the quantiles 
$\Fhat_{Y|X}^{-1}(\cdot|j)$ are plotted against the quantiles of the
empirical distribution of $\{y_i: x_i=j\}$, that is,
the model-based quantiles $\Fhat_{Y|X}^{-1}\left((m - 0.5)/{n_j}\big|
j\right)$ for $m = 1, \ldots, n_j$ are plotted against 
the sorted values of $\{y_i: x_i=j\}$.
If the points in each conditional Q-Q plot lie closely along the $45^{\circ}$ diagonal line, it indicates that the copula estimate fits the sample well. 

The use of these conditional Q-Q plots for copula diagnostics is illustrated in Section~\ref{sec:illus} for simulated datasets and Section~\ref{sec:app} for a real dataset.

\section{Assessing Copula Approximations of Probability Models for Mixed Continuous-Ordinal Variables} 
\label{sec:kl}

For a mix of several continuous variables and one ordinal variable, there are several classes of probability models in the statistical literature.
Since our primary goal is to examine whether simple parametric bivariate copula
families within the vine pair-copula construction are adequate,
we focus on probability models for one continuous and one ordinal variable
and assess the adequacy of approximations by parametric bivariate copula families with a few
parameters.

This section discusses methods to theoretically assess the adequacy of copula approximations by computing the Kullback-Leibler (KL) divergence 
of $C(F_X,F_Y;\theta)$ with a parametric copula family $C(\cdot;\theta)$
relative to a given bivariate distribution $G_{X,Y}$ with margins $F_X$ and $F_Y$.
This leads to summaries of conditions for when parametric copula families with a few parameters are adequate, and other conditions for when they are inadequate. In particular, Section~\ref{subsec:prob_models} gives an overview of common probability models for a pair of mixed continuous-ordinal variables. 
Computational details of the KL divergence 
are covered in Section~\ref{subsec:approximation}. 
Some representative  concrete examples 
are considered in Section~\ref{subsec:klexample}
to show a range of KL divergence values.

\subsection{Probability Models for Mixed Continuous-Ordinal Variables} 
\label{subsec:prob_models}

There are two classes of models for a mix of continuous and ordinal variables depending on the order of conditioning. 

For the first class, a multinomial distribution is assumed for the ordinal
variable and the conditional distribution of the continuous variable
given the ordinal variable can be either Gaussian with different mean and variance parameters, or 
a more general location-scale family (\cite{little1985maximum} and \cite{krzanowski1993location}).
For the second class, the continuous variable is transformed to have
a parametric distribution such as $N(0,1)$ and the ordinal variable conditioned on the
continuous variable is an ordinal probit or logit model;
references are \cite{cox1972analysis} and \cite{krzanowski1993location}.
We introduce the notation for these two classes of models with an ordinal variable $X$ and a continuous variable $Y$.

For the first class of models, a location-scale mixture model is considered. The conditional distribution $F_{Y|X}(\cdot|x)$ is a general location-scale model with
\begin{equation*}
  [Y|X=x] \sim \frac{1}{\sigma_x}p\left(\frac{y - \mu_x}{\sigma_x}\right), 
  \quad x\in\{1,\ldots,k\},
\end{equation*}
where $p$ is a density function on the real line.
The ordinal variable $X$ has a probability distribution
representing the mixture components.
If $Y$ can only take positive values, it can be transformed via a logarithm to get a location-scale model. 

For the second class of models, the conditional distribution 
$F_{X|Y}(\cdot|y)$ is
\begin{equation*}
  \mathbb{P}(X\leq x|Y=y) =F(ay+b_x),\quad x=1,\ldots,k, \quad
  b_1<\cdots<b_{k-1}<b_k=\infty, 
\end{equation*}
where $F$ is a standard normal or logistic CDF, $a$ is a slope parameter, and $b$ is an offset depending on the ordinal category.
The continuous variable $Y$ is assumed to have a unimodal distribution. 
If $Y\sim N(0,1)$, simpler calculation results can be obtained when $F$ is the standard normal CDF.

\subsection{Assessing Copula Approximation via the KL Divergence}
\label{subsec:approximation}

Let $H_{X,Y},G_{X,Y}$ be two bivariate distributions with densities 
$h_{X,Y},g_{X,Y}$ on the respective relevant measure spaces.
For ordinal $X$ with values in $\{1,\ldots,k\}$ and absolutely continuous $Y$,
the product measure comes from a counting measure and the Lebesgue
measure on the real line.
The non-negative KL divergence of $h_{X,Y}$ from $g_{X,Y}$ is
\begin{equation}
 KL\left(h_{X,Y},g_{X,Y}\right)=\int_{-\infty}^\infty \sum_{i=1}^k g_{X,Y}(i,y) \log [g_{X,Y}(i,y)/
  h_{X,Y}(i,y)]\, {\rm d}y.
  \label{eq:KLdiv}
\end{equation}
Let $C(\cdot;\theta)$ be a bivariate copula family with $C_{1|2}(u|v;\theta)={\partial
C(u,v;\theta)/\partial v}$.
If $H_{X,Y}(x,y;\theta)=C(F_X(x),F_Y(y);\theta)$ with conditional probability mass function
$h_{X|Y}(x|y;\theta) = C_{1|2}(F_X(x)|F_Y(y);\theta) - C_{1|2}(F_X(x^-)|F_Y(y);\theta)$,
then the joint density is $h_{X,Y}(x,y;\theta)=f_Y(y)\,h_{X|Y}(x|y;\theta)$ with
$f_Y(y)=\mathrm{d}F_Y(y)/\mathrm{d}y$.
In the setting of a mixture model that specifies $f_X$ and $g_{Y|X}$,
$f_Y(y) = \sum_{i=1}^k f_X(i)\,g_{Y|X}(y|i)$.

The copula with parameter estimate $$\hattheta=\argmin_\theta KL\left(h_{X,Y}(\cdot;\theta),g_{X,Y}\right)$$
is the member in the family that is the closest to $g_{X,Y}$.
The value $KL(h_{X,Y}(\cdot;\hattheta),g_{X,Y})$ is considered as the KL divergence
of the family $\{h_{X,Y}(\cdot;\theta)\}$ from $g_{X,Y}$.

In the results below, we consider different parametric copula families denoted by $C^{(m)}(\cdot;\theta^{(m)})$ with corresponding densities $h^{(m)}_{X,Y}(\cdot; \theta^{(m)})$.
Given $g_{X,Y}$, this leads to $KL(h^{(m)}_{X,Y}(\cdot;\hattheta^{(m)}),g_{X,Y})$
for model $m$. 
A copula family that has a smaller value of KL divergence is considered to have a better approximation of $g_{X,Y}$.

Next are steps to find a sequence of $C^{(m)}$ that leads to smaller
values of the KL divergence of a family from a given $g_{X,Y}$.
Because many simple parametric copula families only have positive
dependence, we assume that $X$ and $Y$
have been oriented to have positive dependence.
We also assume that $Y$ has a distribution that is close to unimodal
(otherwise one might use a non-copula-based model for $(X,Y)$). 
Unless an additional reference is given, properties of the listed copula
families can be found in Chapter 4 of \cite{joe2014dependence}.




\begin{enumerate}
\itemsep=0pt
\item 
Compute the KL divergence of the Gaussian copula family as a baseline.
\item 
For copula family candidates with more dependence in one joint tail,
compute the KL divergence of the Gumbel and survival Gumbel copula families
with asymmetric dependence in the joint upper and lower tails. 
If one of these two families leads to smaller KL divergence, then try copula
families with even more tail asymmetry in the same direction. 
\item
For copula family candidates with less dependence in both joint tails
(compared to Gaussian),
compute the KL divergence of the Frank and Plackett copula families with reflection
symmetric tail quadrant independence.
If one of these two copula families leads to smaller KL divergence,
then compute the KL divergence for the BB8 and BB10 \footnote
{Note that \cite{Kadhem.Nikoloulopoulos2021} show that
there are parameters that can lead to
non-convex contours of copula densities with N(0,1) margins for the BB10 copula.}
families with reflection tail asymmetry and their survival counterparts. 
\item
For copula family candidates with more dependence in both joint tails,
compute the KL divergence for the t copula family.
\item 
If permutation asymmetry is possible in $g_{X,Y}$ (e.g., with heterogeneous
components in a mixture model),
compute the KL divergence for some copula families with permutation asymmetry; examples are skew normal
(\cite{Yoshiba2018}) and asymmetric Gumbel (Section 4.15 of \cite{joe2014dependence}) copula families.
\end{enumerate}

The concept of tail order (\cite{hua.joe2011}), covering copula families
with different strengths of dependence in the joint lower and upper tails,
is used in the above steps. 
The main distinctions are
intermediate tail dependence (such as Gaussian copula with positive dependence),
strong tail dependence (more dependence in the joint tail than Gaussian),
and tail quadrant independence (less dependence in the joint tail than Gaussian).
The concept of tail order may be less important when one variable
is ordinal, because there are no observations in the joint upper or
lower tail. 
However, copula families with tail asymmetries are still important to provide more flexibility when finding approximations to a given $g_{X,Y}$.

\subsection{Examples of KL Divergence for Mixed Continuous-Ordinal Variables} \label{subsec:klexample}

In this section, we consider a variety of concrete examples of the probability models 
in Section \ref{subsec:prob_models} which lead to $g_{X,Y}$ in \eqref{eq:KLdiv}.
In all of these examples, the marginal density $f_Y$ of $g_{X,Y}$
is close to unimodal.
Otherwise, $\text{cor}(X,Y)$ could be large and ``clusters" might be seen
in bivariate scatterplots.
After examining a large number of cases,
we find that a KL
divergence value less than 0.003 usually indicates a good approximation
from a copula family,
and a KL divergence value greater than 0.01 usually indicates
a poor approximation when using the conditional Q-Q diagnostic plots in
Section \ref{subsec:diag}.

The tables in Section~\ref{subsec:cases} summarize some representative
examples to illustrate what happens for (a) mixture models with
different separation of location parameters and common versus varying scale
parameters, and (b) conditional probit or logit models. Section~\ref{subsec:conc_prob_models} contains conclusions drawn from these examples.

\subsubsection{Representative Cases} \label{subsec:cases}

Tables~\ref{tab:2class} and \ref{tab:3class} have some illustrative
examples for 2 and 3 ordinal categories, respectively.
The top parts of these tables have mixture components $[Y|X=x]$ that are Gaussian, t, or skew normal.
The parametric bivariate copula families that lead to the smallest KL divergence, as well as the minimized KL divergence values, are shown in these tables.

For mixture models of different distributions, the mixing proportion vector $\pi$
of $X$ can vary across different cases. 
These examples have unequal mixing proportions and some examples have a dominant component.
In Table~\ref{tab:2class} with two ordinal categories,
models (A1), (B1), and (C1) have close locations and constant scales; 
models (A2), (B2), and (C2) have locations that are farther apart and constant scales; 
models (A3), (B3), and (C3) have close locations and non-constant scales; 
models (A4), (B4), and (C4) have locations that are
farther apart and non-constant scales. 
In Table~\ref{tab:3class} with three ordinal categories,
models (E1), (F1), and (G1) have equally spaced locations and constant scales;  
models (E2), (F2), and (G2) have unequally spaced locations and constant scales;
models (E3), (F3), and (G3) have equally spaced locations and non-constant scales; 
models (E4), (F4), and (G4) have unequally spaced locations and non-constant
scales. 
The tables show that the KL divergence values tend to be smaller for
cases of close or equally spaced locations and constant scales, and
be larger for cases of distant or unequally spaced locations and non-constant scales.
With enough heterogeneity, the conditional Q-Q plots based on the
parametric copula family with the smallest KL divergence typically show
some deviation in the tails.
When the ordinal variable has four or more categories, it is more difficult
for simple parametric copula families to approximate mixture models well. 

If there are asymmetries in the proportions in $f_X$ or the
scale parameters are unequal, the best parametric copulas based on the KL
divergence will have reflection or tail asymmetry, but need not be good fits based on the
conditional Q-Q plots; see Section \ref{sec:illus} for examples of cases (E1) to (E4).
With asymmetries and unobserved extremes of the ordinal variable,
the best parametric copula family based on the KL divergence could have tail dependence in the joint upper and/or lower tail, or in neither joint tail. However, the concept of tail dependence is less
meaningful when one of the variables is ordinal. 

The bottom parts of Tables \ref{tab:2class} and \ref{tab:3class}
have ordinal regression models for the ordinal variable $X$ with
a continuous covariate $Y$. 
For the conditional probit model (D1), $\mathbb{P}(X = 0 | Y = y) = 1 -
\mathbb{P}(Z \leq ay+b) = \mathbb{P} (Z \leq -ay - b)$, where $Z\sim N(0,1)$
and $Z$ is independent from $Y\sim N(0,1)$. 
As a result,  $\mathbb{P}(X = 0) = \mathbb{P}(Z + aY \leq -b)$. 
Let $W = Z + aY$. 
Then $W\sim N(0, 1 + a^2)$ and $\text{cor}(W, Y) = a/(\sqrt{1+a^2})$. 
The binary variable $X$ can thus be considered as being generated from a
latent Gaussian variable $W$ with the cutpoint $-b$. 
Therefore, no matter what values of $a$ and $b$ are used to specify the Bernoulli distributions for the conditional distribution of $X$ given $Y$, a bivariate Gaussian copula with $\rho = a/(\sqrt{1+a^2})$ always provides a perfect match, leading to a KL divergence of 0.

For the conditional logit models (D2) and (D3) when $Y$ has a normal distribution, Gaussian copulas can provide a good approximation since the logit function is very close to the probit function, while t copulas (degrees of freedom 28 and 11) approximate these models slightly better with the KL divergence being very close to 0. 
For the conditional probit or logit models (D4) to (D7) when $Y$ has other
distributions such as t or extreme value (EV) distributions, the
approximation of copula-based models are slightly worse than the previous two scenarios, but still adequate.

\begin{table}[!ht]
  \centering
  \resizebox{\linewidth}{!}{
    \begin{tabular}{*{5}{c}}
      \toprule
      \multicolumn{5}{c}{Mixture of normal distributions} \\\midrule
      Model & $\pi$ & Model parameters & Copula family & KL Divergence \\\midrule
      (A1) & $(0.5, 0.5)$ & $\boldsymbol{\mu} = (1,2), \boldsymbol{\sigma} = (1,1)$ & Gaussian & 0.0001 \\
      (A2) & $(0.7, 0.3)$ & $\boldsymbol{\mu} = (1,3), \boldsymbol{\sigma} = (1, 1)$ & BB8 & 0.0010 \\
      (A3) & $(0.6, 0.4)$ & $\boldsymbol{\mu} = (1,2), \boldsymbol{\sigma} = (1,1.5)$ & Survival BB1 & 0.0040 \\
      (A4) & $(0.2, 0.8)$ & $\boldsymbol{\mu} = (1,3), \boldsymbol{\sigma} = (1, 2)$ & BB1 & 0.0087 \\\midrule
      \multicolumn{5}{c}{Mixture of t distributions ($\nu$: degrees of freedom)} \\\midrule
      Model & $\pi$ & Model parameters & Copula family & KL Divergence \\\midrule
      (B1) & $(0.4, 0.6)$ & $\boldsymbol{\mu} = (1,2), \boldsymbol{\nu} = (3,3)$ & Survival BB10 & 0.0022 \\
      (B2) & $(0.3, 0.7)$ & $\boldsymbol{\mu} = (1,3), \boldsymbol{\nu} = (3,3)$ & Survival BB10 & 0.0057 \\
      (B3) & $(0.6, 0.4)$ & $\boldsymbol{\mu} = (1,2), \boldsymbol{\nu} = (3,6)$ & Survival BB10 & 0.0040 \\
      (B4) & $(0.7, 0.3)$ & $\boldsymbol{\mu} = (1,3), \boldsymbol{\nu} = (3,6)$ & Survival BB10 & 0.0050 \\\midrule
      \multicolumn{5}{c}{Mixture of skew normal distributions ($\alpha$: skew)} \\\midrule
      Model & $\pi$ & Model parameters & Copula family & KL Divergence \\\midrule
      (C1) & $(0.3, 0.7)$ & $\boldsymbol{\mu} = (1,2), \boldsymbol{\sigma} = (1,1), \boldsymbol{\alpha} = (3,3)$ & Survival Joe & 0.0050 \\
      (C2) & $(0.6, 0.4)$ & $\boldsymbol{\mu} = (1,3), \boldsymbol{\sigma} = (1,1), \boldsymbol{\alpha} = (3,3)$ & Survival Joe & 0.0073 \\
      (C3) & $(0.5, 0.5)$ & $\boldsymbol{\mu} = (1,2), \boldsymbol{\sigma} = (1,1.5), \boldsymbol{\alpha} = (3,6)$ & Clayton & 0.0065 \\
      (C4) & $(0.4, 0.6)$ & $\boldsymbol{\mu} = (1,3), \boldsymbol{\sigma} = (1,2), \boldsymbol{\alpha} = (3,6)$ & Survival Joe & 0.0056 \\\midrule
      \multicolumn{5}{c}{Conditional probit or logit models} \\\midrule
      Model & \multicolumn{2}{c}{Model specifications} & Copula family & KL Divergence  \\\midrule
      (D1) & \multicolumn{2}{c}{$Y\sim N(0,1), [X|Y=y]\sim \text{Bernoulli}\left(\Phi(ay + b)\right)$} & Gaussian & 0 \\
      (D2) & \multicolumn{2}{c}{$Y\sim N(0,1), [X|Y=y]\sim \text{Bernoulli}\left(1/\left(1+ \exp(-y - 3)\right)\right)$} & t(28) & $1.5\times 10^{-5}$ \\
      (D3) & \multicolumn{2}{c}{$Y\sim N(0,1), [X|Y=y]\sim \text{Bernoulli}\left(1/\left(1+ \exp(-2y + 2)\right)\right)$} & t(11) & $4.5\times 10^{-5}$ \\
      (D4) & \multicolumn{2}{c}{$Y\sim t_3, [X|Y=y]\sim \text{Bernoulli}\left(\Phi(y + 2)\right)$} & t(8) & 0.0021 \\
      (D5) & \multicolumn{2}{c}{$Y\sim t_3, [X|Y=y]\sim \text{Bernoulli}\left(1/\exp(-y - 1)\right)$} & Gaussian & 0.0021 \\
      (D6) & \multicolumn{2}{c}{$Y\sim \text{EV}, [X|Y=y]\sim \text{Bernoulli}\left(\Phi(y + 1)\right)$} & Gaussian & 0.0019 \\
      (D7) & \multicolumn{2}{c}{$Y\sim \text{EV}, [X|Y=y]\sim \text{Bernoulli}\left(1+ \exp(-y - 1)\right)$} & Gaussian & 0.0011 \\
      \bottomrule
    \end{tabular}
  }
\caption{Bivariate copula families that minimize the KL divergence to probability models for an ordinal variable with two categories and a continuous variable.
For mixture models, $\pi$ is the vector of mixing proportions. 
The minimized KL divergence values are shown in the last column. 
The skew normal distribution has PDF $f(y) = \frac{2}{\sigma}\phi\left(\frac{y-\mu}{\sigma}\right)\Phi\left(\alpha\left(\frac{y-\mu}{\sigma}\right)\right)$ with a skew parameter $\alpha$. The extreme value (EV) distribution has CDF $F(y) = \exp(-\exp(-y)), y\in\mathbb{R}$.}
  \label{tab:2class}
\end{table} 


Results are similar with more than two ordinal categories.
For the conditional ordinal probit model (H1) with three categories 1, 2, and 3, the categorical variable $X$ can be considered as being generated
from a latent Gaussian variable $W$ with cutpoints $b_1$ and $b_2$ with $b_1 < b_2$, where $W\sim N(0, 1+a^2)$ and $\text{cor}(W, Y) = -a/(\sqrt{1+a^2})$. 
No matter what constants $a$, $b_1$, and $b_2$ are used to specify the
probabilities of the three categories, the bivariate Gaussian copula with $\rho = -a/(\sqrt{1+a^2})$ always provides a perfect match, leading to KL divergences of 0. 
The same conclusion extends to conditional ordinal probit models with an arbitrary number of categories when $Y$ follows a normal distribution.
For the conditional logit model (H2) when $Y$ has a normal distribution, the t
copula (degrees of freedom 20) provides a good approximation with KL
divergence being very close to 0. For the conditional probit models (H3) and (H4) when $Y$ has t or EV distributions, the approximation of simple parametric copula families becomes
slightly worse.

\begin{table}[!ht]
  \centering
  \resizebox{\linewidth}{!}{
    \begin{tabular}{*{5}{c}}
      \toprule
      \multicolumn{5}{c}{Mixture of normal distributions} \\\midrule
      Model & $\pi$ & Model parameters & Copula family & KL Divergence \\\midrule
      (E1) & $(0.3, 0.3, 0.4)$ & $\boldsymbol{\mu} = (1,2,3), \boldsymbol{\sigma} = (1,1,1)$ & Gaussian & 0.0016 \\
      (E2) & $(0.5, 0.2, 0.3)$ & $\boldsymbol{\mu} = (1,3,6), \boldsymbol{\sigma} = (2,2,2)$ & Survival BB1 & 0.0031 \\
      (E3) & $(0.4, 0.4, 0.2)$ & $\boldsymbol{\mu} = (1,2,3), \boldsymbol{\sigma} = (3,2,4)$ & Asymmetric Gumbel & 0.0267 \\
      (E4) & $(0.3, 0.4, 0.3)$ & $\boldsymbol{\mu} = (1,3,6), \boldsymbol{\sigma} = (4,6,3)$ & Survival BB1 & 0.0472 \\\midrule
      \multicolumn{5}{c}{Mixture of t distributions ($\nu$: degrees of freedom)} \\\midrule
      Model & $\pi$ & Model parameters & Copula family & KL Divergence \\\midrule
      (F1) & $(0.2, 0.5, 0.3)$ & $\boldsymbol{\mu} = (1,2,3), \boldsymbol{\nu} = (4,4,4)$ & Plackett & 0.0028 \\
      (F2) & $(0.4, 0.2, 0.4)$ & $\boldsymbol{\mu} = (1,3,7), \boldsymbol{\nu} = (4,4,4)$ & Survival BB10 & 0.0432 \\
      (F3) & $(0.4, 0.3, 0.3)$ & $\boldsymbol{\mu} = (1,2,3), \boldsymbol{\nu} = (6,3,9)$ & Survival BB10 & 0.0075 \\
      (F4) & $(0.3, 0.5, 0.2)$ & $\boldsymbol{\mu} = (1,3,7), \boldsymbol{\nu} = (6,3,9)$ & BB8 & 0.0039 \\\midrule
      \multicolumn{5}{c}{Mixture of skew normal distributions ($\alpha$: skew)} \\\midrule
      Model & $\pi$ & Model parameters & Copula family & KL Divergence \\\midrule
      (G1) & $(0.2, 0.4, 0.4)$ & $\boldsymbol{\mu} = (2,3,4), \boldsymbol{\sigma} = (3,3,3), \boldsymbol{\alpha} = (4,4,4)$ & Survival Gumbel & 0.0102 \\
      (G2) & $(0.2, 0.3, 0.5)$ & $\boldsymbol{\mu} = (2,4,8), \boldsymbol{\sigma} = (3,3,3), \boldsymbol{\alpha} = (4,4,4)$ & Survival BB10 & 0.0424  \\
      (G3) & $(0.3, 0.2, 0.5)$ & $\boldsymbol{\mu} = (2,3,4), \boldsymbol{\sigma} = (3,1,2), \boldsymbol{\alpha} = (3,2,4)$ & t(2) & 0.1421 \\
      (G4) & $(0.5, 0.3, 0.2)$ & $\boldsymbol{\mu} = (2,4,8), \boldsymbol{\sigma} = (3,1,2), \boldsymbol{\alpha} = (3,2,4)$ & BB8 & 0.2540 \\\midrule
      \multicolumn{5}{c}{Conditional probit or logit models} \\\midrule
      Model & \multicolumn{2}{c}{Model specifications} & Copula family & KL Divergence  \\\midrule
      (H1) & \multicolumn{2}{c}{$Y\sim N(0,1), \begin{cases}
        \mathbb{P}(X\leq1|Y=y) = \Phi(ay + b_1)\\
        \mathbb{P}(X\leq2|Y=y) = \Phi(ay + b_2),\end{cases}$} $b_1 < b_2$  & Gaussian & 0 \\\hdashline
      (H2) & \multicolumn{2}{c}{$Y\sim N(0,1), \begin{cases}
      	\mathbb{P}(X\leq1|Y=y) = 1/\exp(y + 1)\\
      	\mathbb{P}(X\leq2|Y=y) = 1/\exp(y - 1)\end{cases}$}  & t(20) & $1.6\times 10^{-5}$ \\\hdashline
      (H3) & \multicolumn{2}{c}{$Y\sim t_3, \begin{cases}
      	\mathbb{P}(X\leq1|Y=y) = \Phi(-y - 1)\\
      	\mathbb{P}(X\leq2|Y=y) = \Phi(-y + 1)\end{cases}$}  & Gaussian & 0.0038 \\\hdashline
      (H4) & \multicolumn{2}{c}{$Y\sim \text{EV}, \begin{cases}
      	\mathbb{P}(X\leq1|Y=y) = \Phi(-y - 1)\\
      	\mathbb{P}(X\leq2|Y=y) = \Phi(-y + 1)\end{cases}$}  & Gaussian & 0.0095 \\
      \bottomrule
    \end{tabular}
  }
\caption{Bivariate copula families that minimize the KL divergence to probability models for an ordinal variable with three categories and a continuous variable. 
The minimized KL divergence values are shown in the last column. 
Other definitions are the same as in Table~\ref{tab:2class}.}
  \label{tab:3class}
\end{table} 


\subsubsection{Conclusions on Copula Approximations to Models in
Section~\ref{subsec:prob_models}} \label{subsec:conc_prob_models}

Based on the representative examples in the previous section, the following
general conclusions can be drawn. Simple parametric copula-based models provide good approximations mainly
for (a) mixture models $[Y|X=x]$ with roughly equally spaced components and similar scale parameters, and (b) ordinal regression models $[X|Y=y]$ that are closed to probit models with a unimodal
distribution for $Y$.
For mixture models with components that have unequally spaced location parameters or 
quite different scale parameters, the effective number of parameters is
greater than 2, so it is not surprising that the simple parametric
copula families do not lead to good approximations.
This motivates the use of nonparametric copulas in Section~\ref{subsec:model2}.

\section{Fitting Copula Models to a Mixed Continuous-Ordinal Pair}
\label{sec:fitting}

This section explains the procedures for fitting copula models to a pair of mixed continuous-ordinal variables. 
For a random sample $\{(x_i,y_i): i=1,\ldots,n\}$,
$u^+_{iX}, u^-_{iX}, u_{iY}, u_{iZ}$ are as defined in Section~\ref{sec:diagnostics}.
Details for fitting parametric and nonparametric copula models are elaborated in Sections~\ref{subsec:model}~and~\ref{subsec:model2}, respectively. 

\subsection{Parametric Bivariate Copula Families} 
\label{subsec:model}

Suppose there are $M$ parametric bivariate copula models to consider as candidate families. 
For a parametric bivariate copula model $C^{(m)}$ with parameter vector $\bftheta^{(m)}$ and   $C_{1|2}^{(m)}(u|v;\bftheta^{(m)})=
\partial C^{(m)}(u,v;\bftheta^{(m)})/\partial v$, its log-likelihood function is
$$
\mathcal{L}_m\bigl(\bftheta^{(m)}\bigr) =
\sum_{i=1}^{n}\log\left\{C_{1|2}^{(m)}\bigl(u_{iX}^+|
u_{iY};\bftheta^{(m)}\bigr) - C_{1|2}^{(m)}\bigl(u_{iX}^-|
u_{iY};\bftheta^{(m)}\bigr)\right\}.
$$
The maximum likelihood estimator is $\widehat{\bftheta}^{(m)}_n =
\argmax_{\bftheta^{(m)}}\mathcal{L}_m(\bftheta^{(m)})$ for a sample of size $n$. 
As $n\to\infty$, $\widehat{\bftheta}^{(m)}_n$ converges in probability to the 
value $\widetilde{\bftheta}^{(m)}$ that minimizes the KL divergence
for family $C^{(m)}(\cdot;\bftheta^{(m)})$.
Parametric copula models are then compared using model selection criteria such as AIC and BIC. 
Models with smaller values of AIC or BIC are usually considered to fit the data better. 

\subsection{Nonparametric Bivariate Copulas} 
\label{subsec:model2}


Since \eqref{eq:condcdf} only involves the copula CDF but not the copula density, we consider the empirical beta copula (\cite{segers2017empirical}) as a nonparametric alternative fitted to
pairs $(u_{iZ},u_{iY})$, $i=1,\ldots,n$.
The empirical beta copula density performs less well than the KDE copula estimator in \cite{nagler2018kdecopula}, even after some averaging over distinct subsamples. 
However, the empirical beta copula CDF has the advantage of being a proper copula, while the KDE approach only leads to a distribution that is approximately a copula.

Let $\bfu_{Z} = \left\{u_{iZ}: i = 1,\dots,n\right\}$ and $\bfu_{Y}  = \left\{u_{iY}: i = 1,\dots,n\right\}$. 
For $\{(u_{iZ}, u_{iY})\}$, the bivariate empirical beta copula CDF is given by
\begin{equation}
C_{n,Z,Y}^{\beta}(u_Z, u_Y) = \frac{1}{n}\sum_{i=1}^{n}F_B\left(u_Z;R^{(n)}_{iZ}, n+1-R_{iZ}^{(n)}\right)\cdot F_B\left(u_Y;R^{(n)}_{iY}, n+1-R_{iY}^{(n)}\right),
 \label{eq: empbeta}
\end{equation}
where $F_B(\cdot;\alpha,\beta)$ is the CDF of the $\text{Beta}(\alpha,\beta)$ distribution, $R_{iZ}^{(n)}$ is the rank of $u_{iZ}$ in $\bfu_{Z}$, and $R_{iY}^{(n)}$ is the rank of $u_{iY}$ in $\bfu_{Y}$.
Note that \eqref{eq: empbeta} is a continuous differentiable function on $[0,1]^2$.
The $r$th order statistic $U_{(r:n)}$ in a random sample of size $n$ generated from $U(0,1)$ follows a $\text{Beta}(r,n+1-r)$ distribution, which leads to the two beta distributions in \eqref{eq: empbeta}. 

Assuming a consistent estimator for $F_Y$,
the consistency of the empirical beta copula with the latent vector $\bfu_Z$
is shown in Section~\ref{subsec:empcop} below. 

\subsection{Consistency of the Empirical Beta Copula Estimate with a Latent Variable} 
\label{subsec:empcop}


Let $F_{n,X}$ be the empirical distribution of the ordinal variable $X$. 
Let $F_{n,Y}$ be a consistent estimate of $F_Y$.
Let $C_{n,Z,Y}^\beta(u_{Z}, u_{Y})$ be the empirical beta copula estimate in \eqref{eq: empbeta}
and let $C_{n,Z,Y}(u_{Z}, u_{Y})$ be the empirical copula of the same sample, defined as
 $$
C_{n,Z,Y}(u_{Z}, u_{Y}) = \frac{1}{n}\sum_{i=1}^{n}\mathbb{I}\left\{\frac{R_{iZ}^{(n)}}{n} \leq u_Z\right\} \mathbb{I}\left\{\frac{R_{iY}^{(n)}}{n} \leq u_Y\right\}, \quad 0\leq u_Z \leq 1, \quad 0\leq u_Y\leq 1.
 $$
The empirical copula is a step function.

Proposition 2.8 in \cite{segers2017empirical} states that 
$$
\sup_{(u_{Z}, u_{Y})\in[0,1]^2}\left|C_{n,Z,Y}(u_{Z}, u_{Y}) -
C_{n,Z,Y}^\beta(u_{Z}, u_{Y}) \right| = O\left(n^{-1/2}(\log n)^{1/2}\right).
$$
This indicates that 
$C_{n,Z,Y}(u_{Z}, u_{Y}) - C_{n,Z,Y}^\beta(u_{Z}, u_{Y}) \to 0$ for arbitrary $u_{Z}$ and $u_{Y}$ as $n\to\infty$. 
It is shown in Section~\ref{subsec:normal_scores} that $u_{iZ}$ can be considered as being sampled from a uniform distribution that satisfies the condition \eqref{eq: Z_cond} as $n\to\infty$. 
Based on results on the empirical copula processes (\cite{Segers2012}),
$C_{n,Z,Y}$ converges weakly to $C_{ZY}$, where $Z$ is the latent Gaussian variable for $X$ in Section~\ref{subsec:normal_scores}. 
Since $C_{ZY}$ matches $C_{XY}$ at the CDF values of $X$ when $Z$ satisfies the condition~\eqref{eq: Z_cond}, $C^\beta_{n,Z,Y}(F_X(i), u_{Y}) - C_{X,Y}(F_X(i), u_{Y}) \overset{p}{\to}0$ for $i = 1,\dots,k$ and arbitrary $u_{Y}$ as $n\to\infty$. 
This shows the consistency of the empirical beta copula estimate in
\eqref{eq: empbeta}.

\section{Illustrations on Simulated Datasets} 
\label{sec:illus}

In this section, we illustrate the visualization, estimation, and diagnostic techniques proposed in the previous sections on simulated datasets.
Bivariate datasets are generated from four mixture models of normal distributions, denoted by (E1), (E2), (E3), and (E4) in Table~\ref{tab:3class}.
In each case, the sample size is 1000.

\begin{figure}[!ht] 
  \centering
  \includegraphics[width=\linewidth]{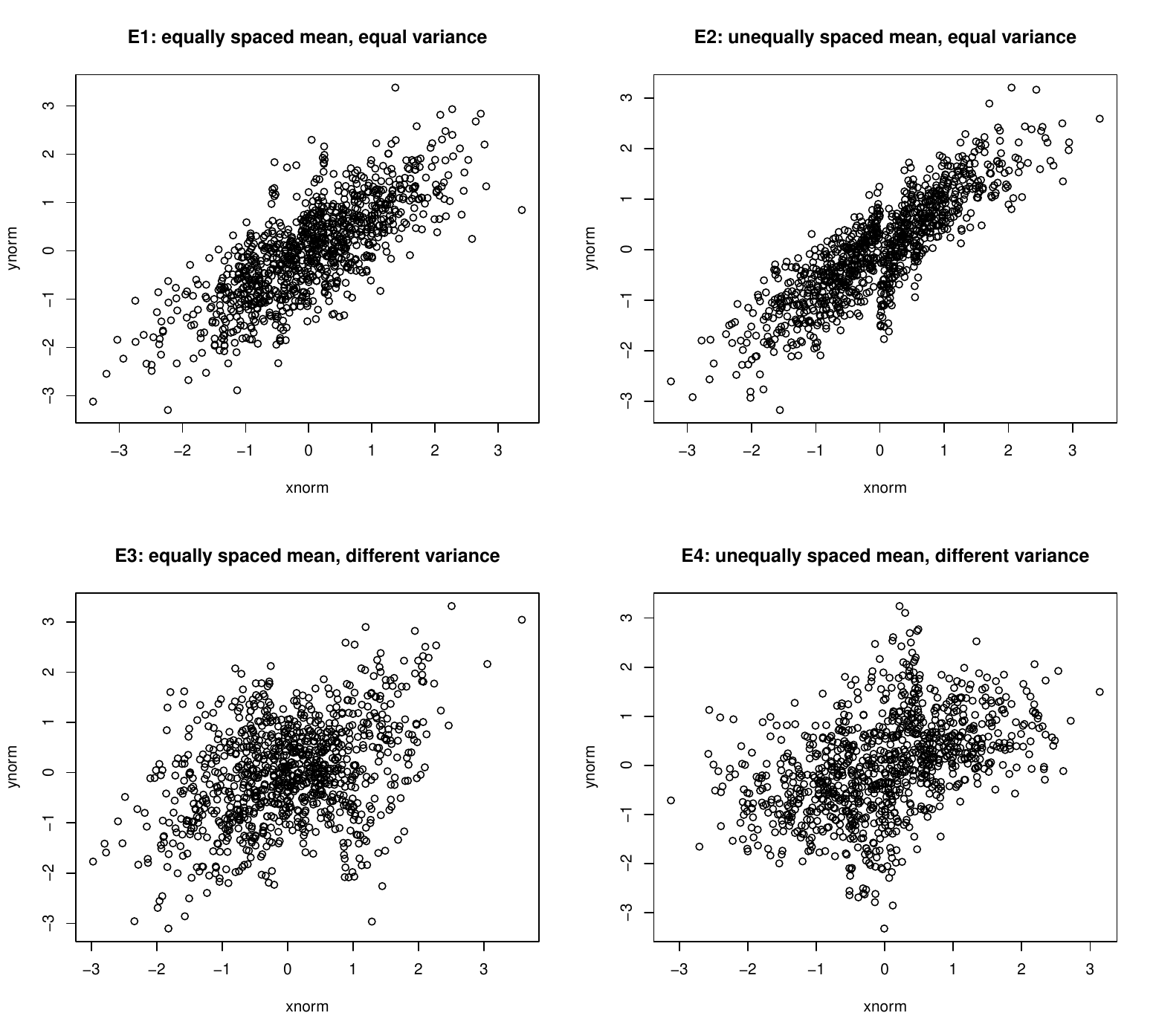}
  \caption{Normal score plots of the continuous variable $Y$ and the latent Gaussian variable $Z$ generated from the ordinal variable $X$ for cases (E1) to (E4).}
  \label{fig: nscore_plot}
\end{figure}

In Table~\ref{tab:3class}, the minimized KL divergence values for cases (E1)
and (E2) are much smaller than those for cases (E3) and (E4), indicating
that parametric copula families fit (E1) and (E2) better than (E3) and (E4). 
Normal score plots of the continuous variable $Y$ versus the latent Gaussian variable $Z$ generated from the ordinal variable $X$ based on the steps in Section~\ref{subsec:normal_scores} are shown in Figure~\ref{fig: nscore_plot}. 
It can be seen that the normal score plots of (E1) and (E2) have an
approximate elliptical shape that can match some commonly used parametric copula families. 
In contrast, for (E3) and (E4), heteroscedasticity among different components of the mixture model leads to asymmetries and unusual shapes in the normal score plots. 
Simple parametric copula families are inadequate for the data generated in these two cases.

\begin{figure}[!ht] 
  \centering
  \includegraphics[width=\linewidth]{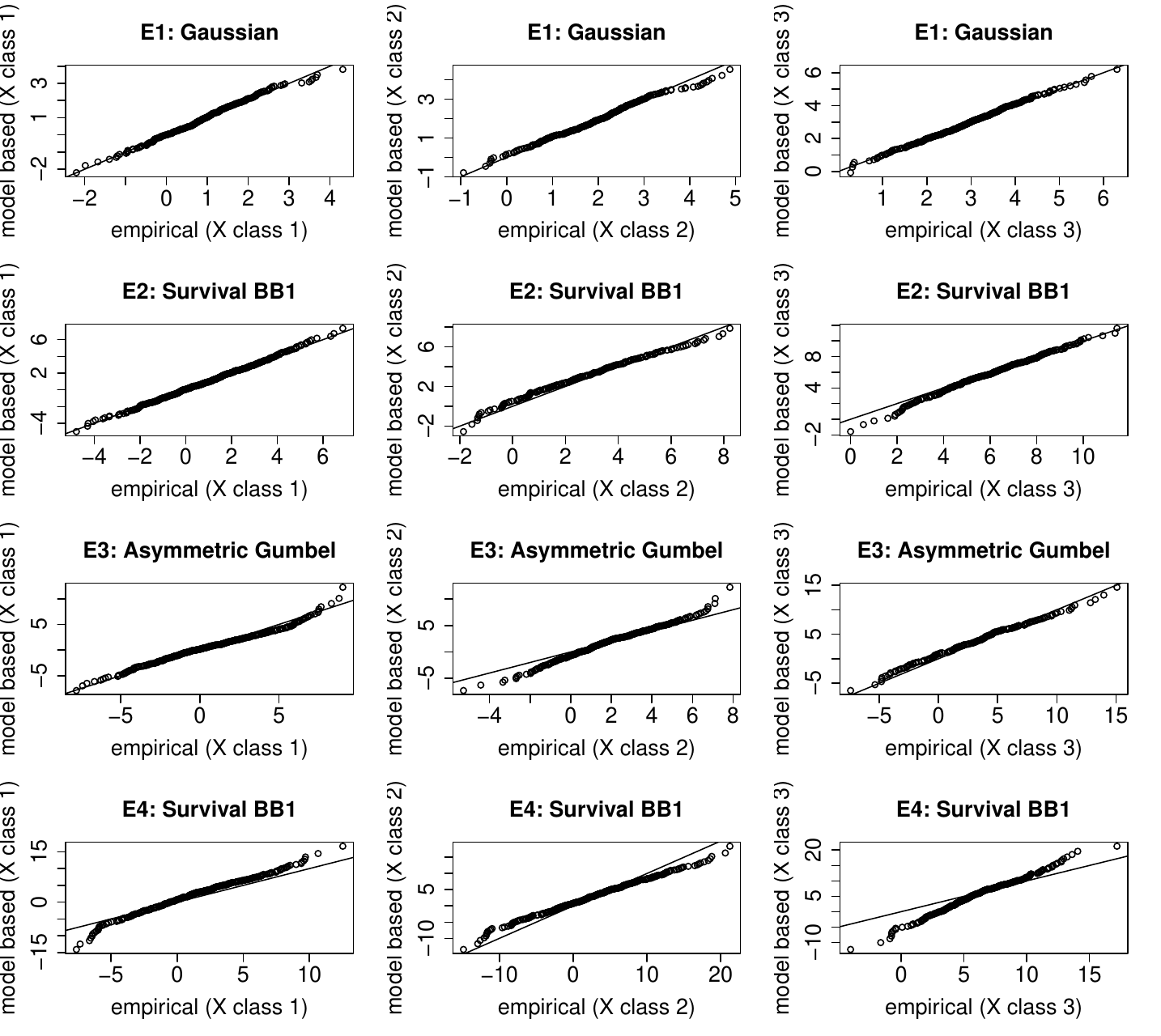}
\caption{The conditional Q-Q plots of $F_{Y|X}(\cdot|x)$ for different
ordinal categories $x$ assessing the fits of the parametric copula families for four simulated datasets based on (E1)--(E4) in Table \ref{tab:3class} with 3 categories. 
Each row corresponds to the conditional Q-Q plots for one dataset.}
  \label{fig: parametric_qq}
\end{figure}

Diagnostic techniques in Section~\ref{subsec:diag} are applied to assess the fits of the parametric copula families in Table~\ref{tab:3class} with the smallest KL divergence.
The conditional Q-Q plots by category of the ordinal variable $X$ are shown in Figure~\ref{fig: parametric_qq}. 
There are significant departures from the $45^\circ$ diagonal line for the bivariate copula families fitted to cases (E3) and (E4), indicating inadequate fits. 
This aligns with the conclusions drawn from the normal score visualizations in Figure~\ref{fig: nscore_plot}. 

\begin{figure}[!ht] 
  \centering
  \includegraphics[width=\linewidth]{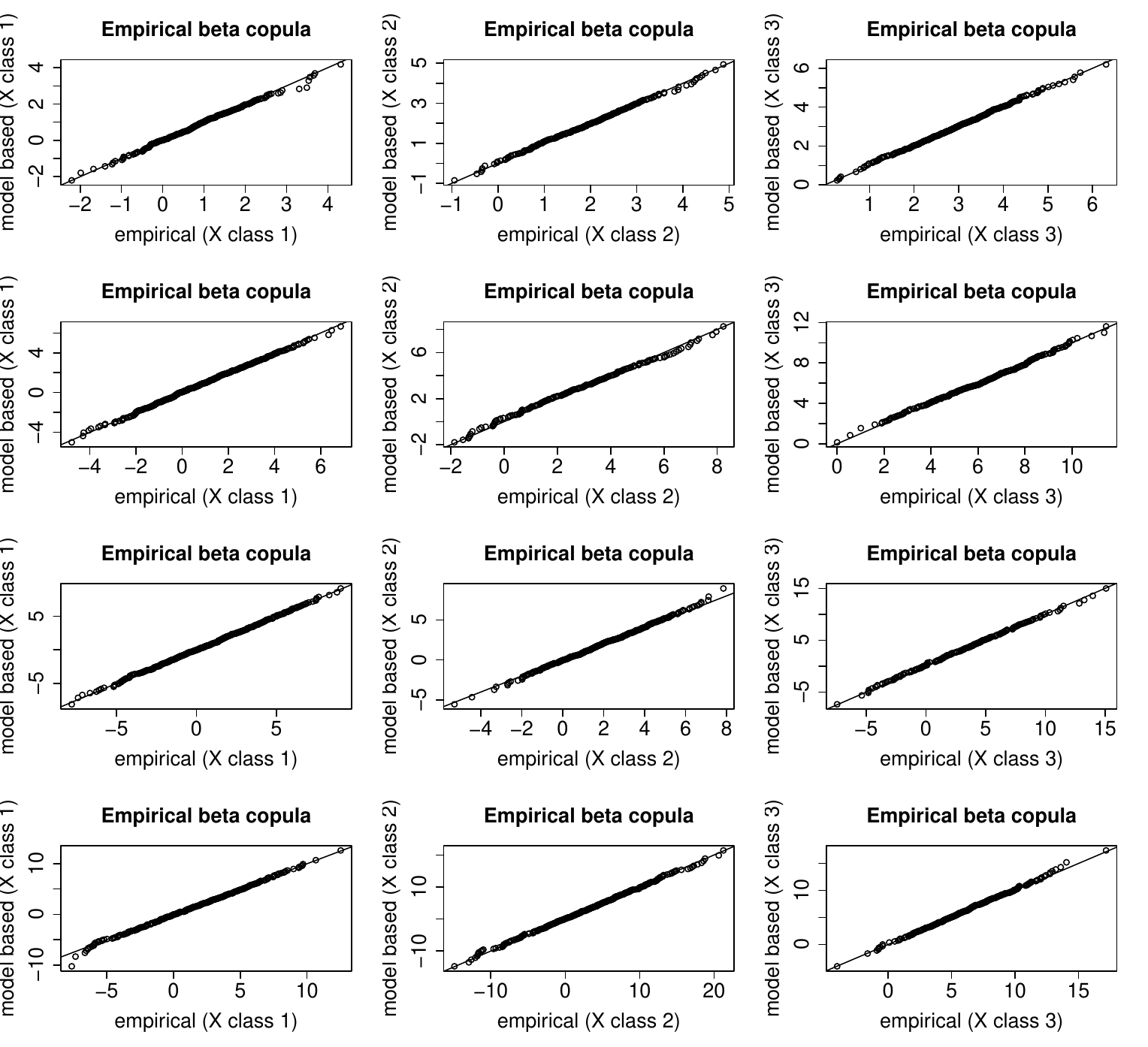}
\caption{
The conditional Q-Q plots of $F_{Y|X}(\cdot|x)$ for different
ordinal categories $x$ assessing the fits of the empirical beta copulas for the four simulated datasets used in Figure~\ref{fig: parametric_qq}.}
\label{fig: nonparametric_qq}
\end{figure}

With the empirical beta copula estimate 
in Section~\ref{subsec:model2}, 
the conditional Q-Q plots by category based on the ordinal variable $X$ are shown in Figure~\ref{fig: nonparametric_qq}. 
It can be seen that the points in all Q-Q plots are closely aligned with the $45^\circ$ diagonal line, indicating that the empirical beta copula estimate provides a much more adequate fit to the data generated in all of these four cases.

\section{Data Example}
\label{sec:app}

In this section, we illustrate the proposed visualization, estimation, and diagnostic techniques for a pair of mixed continuous-ordinal variables on the Auto MPG dataset (\cite{quinlan1993combining}, available at \url{https://archive.ics.uci.edu/ml/datasets/Auto+MPG}).

This dataset contains the fuel consumption data of 398 cars from 1970 to 1982. 
The goal is to predict the \textit{mpg} (miles per gallon as an
indicator of fuel efficiency) of each car based on a set of explanatory variables. 
The variable \textit{cylinders} can take five unique ordinal values: 3, 4, 5, 6, and 8. 
We merge category 3 with 4 and category 5 with 6 since only 4 and 3 observations have 3 and 5 cylinders. 
Some summaries of the important variables along with the transformations to achieve positive correlations with the response variable \textit{mpg} are given in Table~\ref{tab:univariate}. 
The nominal variable \textit{origin} indicates where the car is from (1: USA, 2: Europe, 3: Japan). 
Since \textit{mpg} tends to increase as \textit{origin} changes from USA to Europe to Japan, we treat \textit{origin} as an ordinal variable for the vine structure selection and conditional Q-Q plots mentioned below.

\begin{table}[!ht]
\centering
\begin{tabular}{lllll}
  \toprule
  Variable & Type & $\rho$ & Range & Transformation \\\midrule
  \textit{cylinders} & ordinal & -0.781 & $\{4,6,8\}$ & change sign, $x=-x$\\
  \textit{horsepower} & continuous & -0.771 & $[46,230]$ & change sign, $x=-x$\\
  \textit{weight} & continuous & -0.832 & $[1613,5140]$ & change sign, $x=-x$\\
  \textit{acceleration} & continuous & 0.420 & $[8.0,24.8]$ &\\
  \textit{model year} & continuous & 0.579 & $[70,82]$ & \\
  \textit{origin} & nominal & 0.563 & $\{1,2,3\}$ &\\
  \textit{mpg} & continuous & & $[9.0, 46.6]$ & \\
  \bottomrule
\end{tabular}
\caption{Summaries of the variables in the Auto MPG dataset, including their type, Spearman's $\rho$ values with the response (\textit{mpg}), range, and transformations to achieve positive correlation with \textit{mpg}.}
\label{tab:univariate}
\end{table}

If the maximum spanning tree criterion based on absolute 
$\rho_N$ (correlation of normal scores, or polyserial/polychoric correlation) as in \cite{chang2019prediction} is used to select the vine structure for this dataset, 
the first tree of the resulting vine regression model would
include the two ordinal variables \textit{cylinders} and \textit{origin}
connected to \textit{weight} on two edges for explanatory variables.
The summary statistics of the distribution of \textit{weight} (before sign change), the conditional distribution of \textit{weight} given \textit{cylinders} (before sign change), and the conditional distribution of \textit{weight} given \textit{origin} are shown in Table~\ref{tab:conditional}. 
Note that when fitting the vine copula model, the signs of \textit{weight} and \textit{cylinders} are changed so that the correlations between \textit{negative weight}, \textit{negative cylinders}, and \textit{origin} are all positive. 
The conditional distributions of \textit{weight} given \textit{cylinders} have
roughly equally spaced means and similar standard deviations; this is close to the setting of a mixture model with equally spaced locations and constant variance. 
Therefore, parametric copula families can fit this pair well. 
In contrast, the conditional distributions of \textit{weight} given
\textit{origin} have unequally spaced means and very different standard
deviations; this is similar to a mixture model with unequally spaced locations and non-constant variance. 
Therefore, it is more difficult for parametric copula families to provide
good approximations to this second pair.

\begin{table}[!ht]
	\centering
	\begin{tabular}{lllllllll}
		\toprule
Subset & $n$ & Min & Q1 & Median & Mean & Q3 & Max & SD \\\toprule
		None & 198 & 1613  &  2224  &  2804  &  2970  &  3608  &  5140 & 847 \\\midrule
		$\textit{cylinders} = 4$ & 208 & 1613  &  2049  &  2240  &  2310  &  2567  &  3270 & 345 \\
		$\textit{cylinders} = 6$ & 87 & 2472  &  2938  &  3193  &  3195  &  3431 &   3907  & 332 \\
		$\textit{cylinders} = 8$ & 103 & 3086   & 3799 &   4140 &   4115  &  4404  &  5140  & 449 \\\midrule
		$\textit{origin} = 1$ & 249 & 1800  &  2720  &  3365  &  3362  &  4054   & 5140 & 795\\
		$\textit{origin} = 2$ & 70 & 1825  &  2067  &  2240  &  2423  &  2770  &  3820 & 490 \\
		$\textit{origin} = 3$ & 79 & 1613  &  1985  &  2155  &  2221  &  2412  &  2930 & 320 \\
		\bottomrule
	\end{tabular}
\caption{The summary statistics, including sample size ($n$), quartiles, mean, and standard
deviation (SD), of the distribution of \textit{weight},
the conditional distributions of \textit{weight} given \textit{cylinders}, and the conditional distributions of \textit{weight} given \textit{origin}.}
\label{tab:conditional}
\end{table}

\begin{figure}[!ht] 
\centering
\includegraphics[width=\linewidth]{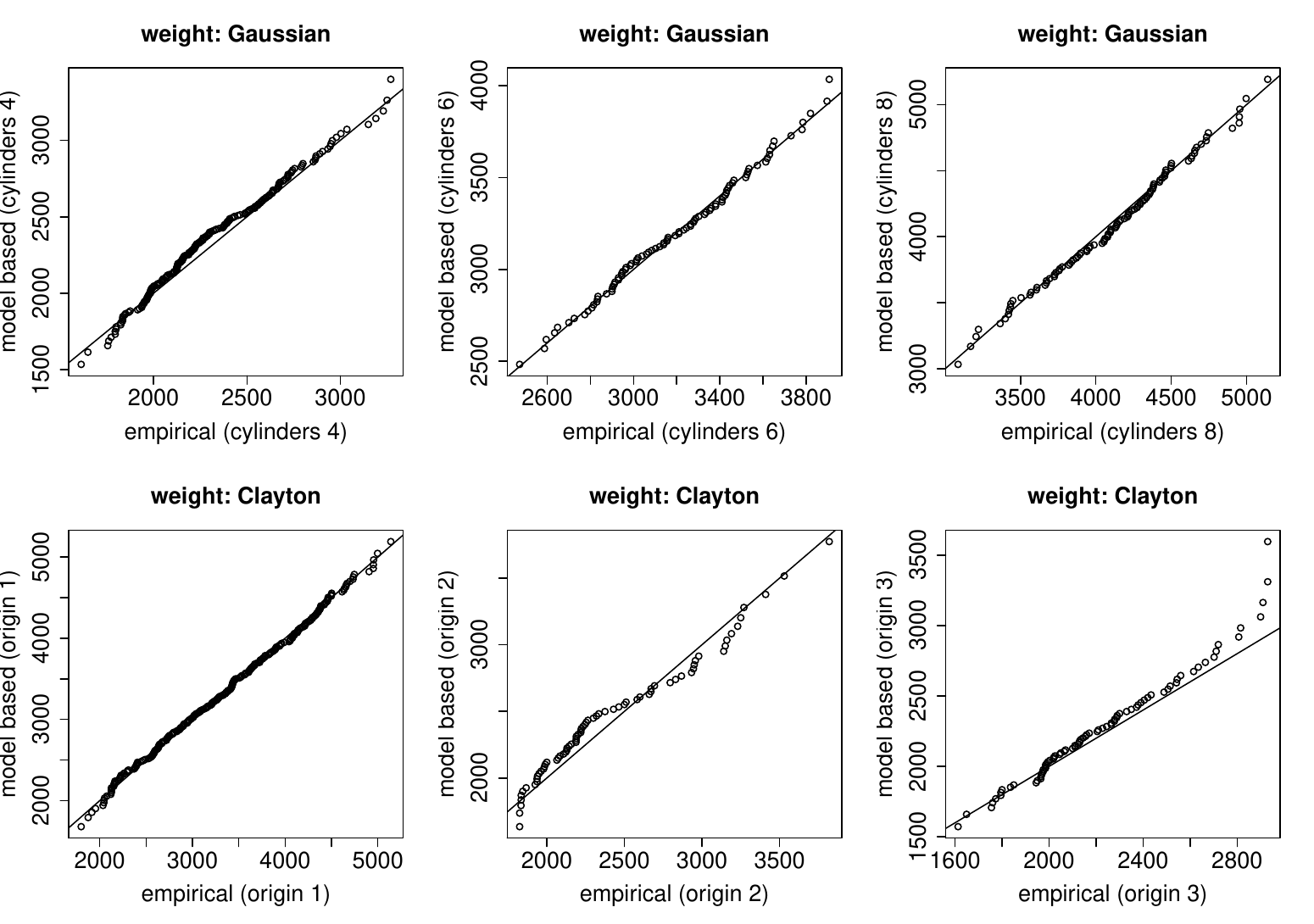}
\caption{The conditional Q-Q plots of $F_{Y|X}(\cdot|x)$ for $Y=$\textit{weight} by category of the ordinal variable $X=$\textit{cylinders} (first row) and the ordinal variable $X=$\textit{origin} (second row) assessing fits of the best-fitting 1-parameter and 2-parameter copula families.}
\label{fig: parametric_qq_5d}
\end{figure}

The best-fitting parametric copula family 
for the pair \textit{negative weight} and \textit{negative cylinders} 
is Gaussian ($\rho = 0.97$).
Conditional Q-Q plots for \textit{weight} by category of \textit{cylinders}
are show in the first row of Figure~\ref{fig: parametric_qq_5d}.
For the pair of \textit{negative weight} and \textit{origin}, the large proportion of the $\textit{origin}=1$ category causes the best-fitting parametric copula families to have tail asymmetry with more dependence in the joint lower tail. 
The 1-parameter Clayton, 2-parameter BB1 and 2-parameter BB7 copulas have the largest
(and approximately equal) log-likelihoods and similar conditional Q-Q plots. 
The conditional Q-Q plots of \textit{weight} by category of \textit{origin} are
shown in the second row of Figure~\ref{fig: parametric_qq_5d} with the Clayton copula. 
The two parametric copula families in this figure generally fit the data well. 
Some deviations from the $45^\circ$ diagonal line can still be observed in
the last plot for the \textit{weight} and \textit{origin} pair, but 
inference based on this copula should be acceptable. 
When the empirical beta copulas are fitted to these two pairs of variables,
all conditional Q-Q plots show alignment with the $45^\circ$ diagonal line, including the last plot.

\section{Conclusion} \label{sec:conc}

Through two types of diagnostic plots and theoretical assessments using the
Kullback-Leibler divergence, we show that simple parametric bivariate copula
families with a few parameters can sometimes be inadequate for a pair of mixed continuous-ordinal variables. 
For such a pair, visualizations are proposed based on
normal score plots using an appropriate latent continuous variable and conditional Q-Q plots of the continuous variable given the ordinal variable. 
Existing probability models for mixed continuous-ordinal variables are considered to assess the adequacy of fits of simple parametric copula families using the Kullback-Leibler divergence. 
When a pair of mixed continuous-ordinal variables is generated from mixture models of distributions with roughly equally spaced locations and constant scales or from conditional probit/logit models, simple parametric copula families can provide good fits. 
Otherwise, nonparametric counterparts can be fitted to provide better approximations.
Applications to simulated and real datasets demonstrate the effectiveness of
the proposed methods in identifying the lack of fit of simple parametric copula families and in improving the adequacy of fits with nonparametric copulas. 


The results in this paper can be used to understand when and how some standard
regression methods with ordinal and continuous explanatory variables can be
approximated by the vine copula regression methodology as considered in \cite{chang2019prediction}. 
Details will be provided in future research.

\section*{Acknowledgments}

This research has been supported by the Four-Year Doctoral Fellowship of the University of British Columbia, NSERC Discovery Grant GR010293, and a Mercator Fellowship associated with Deutsche Forschungsgemeinschaft.

\bibliographystyle{apalike}
\bibliography{literature}

\end{document}